# Chemical gas sensor based on a novel capacitive microwave flexible transducer and composite polymer carbon nanomaterials

P. Bahoumina[1], H. Hallil[1], J.L. Lachaud[1], D. Rebière[1], C. Dejous[1]
[1] Univ. Bordeaux, Bordeaux INP, CNRS, IMS UMR 5218
F-33405 Talence, France
prince.bahoumina@ims-bordeaux.fr

A. Abdelghani[2], K. Frigui[2], S. Bila[2], D. Baillargeat[2]
[2] Univ. Limoges/CNRS, XLIM UMR 7252
F-87060 Limoges

Q. Zhang[3], P. Coquet[3,4]
[3] CINTRA, CNRS/NTU/THALES, UMI 3288
Singapore 637553, Singapore

C. Paragua[4], E. Pichonat[4], H. Happy[4]
[4] Univ. Lille, CNRS, IEMN UMR 8520
F-59652 Villeneuve d'Ascq, France

*Abstract*—This study presents the results on the feasibility of a resonant planar chemical capacitive sensor in the microwave frequency range suitable for gas detection and for wireless communications applications. The objective is to develop a low cost ultra-sensitive sensor that can be integrated into a real time multi-sensing platform. The first demonstrators target the detection of harmful gases such as volatile organic compounds (VOCs) to monitor environmental pollution.

*Keywords— chemical gas sensor; electromagnetic transduction; inkjet printing; flexible substrate; carbon materials; microwave device; resonator*

## I. INTRODUCTION

The environmental pollution caused 7 million deaths worldwide in 2012 according to the World Health Organization [1] and it mobilizes every year increasing resources for control and monitoring. Such issue justifies the need for autonomous, highly selective and sensitive sensors at low cost and low energy consumption. In this context and to contribute to this major societal challenge, we propose a real-time monitoring and quantification of toxic compounds by combining a resonant passive electromagnetic transducer with a sensitive layer based on polymer carbon nanocomposites. This component operates in the frequency range up to 6 GHz; it aims at a suitable solution for the Internet of Things and embedded systems, offering interesting prospects for the proliferation of detection and control communicating wireless networks sites. In addition, the device can be fabricated on a flexible substrate by low cost printing technology [2], [3], [4]. We present here the theoretical principle and the design of two prototypes each consisting of two bandpass resonators on paper. Results of simulation, electrical characterization and under gas are proposed and analyzed, before concluding and opening on perspectives.

## II. THEORETICAL STUDY

### A. Design of sensors

The devices consist of two bandpass resonators on a flexible paper substrate in order to achieve differential detection. A resonator without any sensitive layer is considered as a reference channel, while the other resonator is the sensitive or measurement channel functionalized with a sensitive material to the target gas. Each channel consists of two parallel networks of 50 electrodes each. The geometry and configuration of these two prototypes $D_1$ and $D_2$ are shown on FIG. 1, with a main difference related to the width of the gap ($L_2$) between two successive electrodes. Indeed, the initially intended gap of 200 μm for $D_1$ has been increased to 300 μm for $D_2$ in order to increase the sensitive area and to facilitate fabrication.

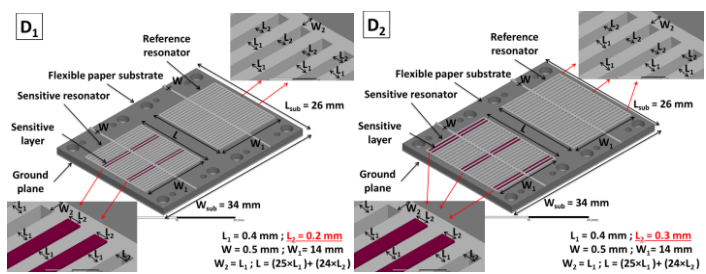

Fig. 1. Representation of the geometry of the devices $D_1$ and $D_2$.

### B. Operating principle and simulations

The operating principle of each sensitive resonator is based on the perturbation of the electromagnetic field due to the presence of the target species on the sensitive layer which causes the variation of its electromagnetic properties, in particular: permittivity, conductivity and / or permeability [2], [3], [4]. The reference channel compensates the variations not related to the sensitive material. In this study, we use a paper substrate with a permittivity of 3.05 and a dielectric loss of 0.089 at 4.3 GHz. An ink based on silver nanoparticles is used for line metallization, a conductive ink of poly (3,4-

ethylenedioxythiophene) polystyrene sulfonate - multi walled carbon nanotubes (PEDOT: PSS - MWCNTs) is used as sensitive layer. The material parameters were previously measured and presented in [5] and [6]. FIG. 2 shows the S parameters resulting from simulation of each device, for their two resonant modes in the frequency range from 1 to 6 GHz. The first and second modes of the $D_1$ reference resonator are recorded at 2.87 and 5.5 GHz with a quality factor Q_$D_1$_Ref equal to 4.1 and 8.6 at -3 dB from the minimum of insertion losses, respectively. Similarly, the first and second modes of the reference resonator of $D_2$ are close to 2.5 and 4.85 GHz with a quality factor Q_$D_2$_Ref equal to 2.79 and 4.71, respectively. The addition of 10 sensitive layers on the sensitive channel of $D_1$ induces a frequency shift equal to +130 MHz and +350 MHz on both modes, respectively. Since the $D_2$ resonator surface is larger than of $D_1$ (300 μm gap), only 5 layers were considered on the corresponding sensitive channel, resulting in a frequency shift close to -100 MHz for each mode.

sensitive layers each were deposited in the 200 μm gaps of the two zones where the E field of the 2nd mode is less intense, in order to favor the interactions with the H field and to have an additional inductive disturbance effect in the presence of gas. On the measurement channel of $D_2$, 12 bands of 5 sensitive layers each were deposited in the 300 μm gaps of the 3 zones where the E field of the 2nd mode is more intense, in order to emphasize the interactions with the E field and so, a capacitive disturbance effect. This can be seen on experimental prototypes (FIG. 5). On each sensitive channel of both devices, we have simulated the presence of 5 to 50 layers of PEDOT: PSS [7], which have similar properties as the PEDOT: PSS - MWCNTs sensitive material used in this study, due to very low concentration of MWCNT.

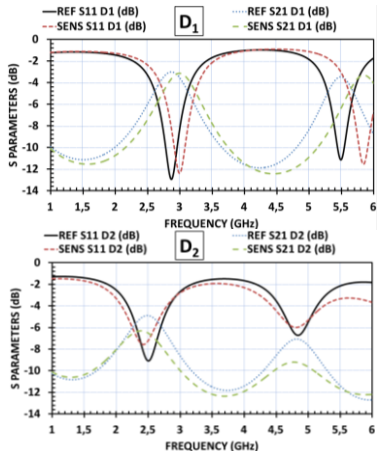

Fig. 2. Simulated S parameters of $D_1$ and $D_2$.

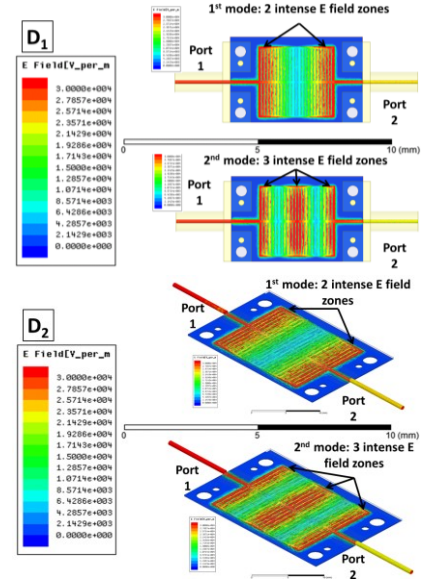

Fig. 3. Distribution of the electric field E for the two resonant modes of $D_1$ and $D_2$.

## C. Analytical model

Analytically, the resonant frequency of each mode of the reference resonator ($f_{rn}$) can be estimated by the relation (1). The effective permittivity ($\varepsilon_{eff}$) can be calculated with expression (2), if the width of the metallic lines is greater than the substrate thickness and considering the same value for the width of the electrodes ($L_1$) and that of the feed lines (W).

$$f_{rn} = \frac{n \times c}{2 \times (W1 + L - W) \times \sqrt{\varepsilon_{eff}}} \quad (1)$$

$$\varepsilon_{eff} = \frac{\varepsilon_r + 1}{2} + \frac{\varepsilon_r - 1}{2}(1 + 12\frac{h}{w_{eff}})^{-\frac{1}{2}} \quad (2)$$

With n ≥ 1, c represents the free space velocity ($3.10^8$ m/s), $\varepsilon_{eff}$ is the effective permittivity of the homogeneous medium, h is the substrate thickness, $\varepsilon_r$ is its relative permittivity and $w_{eff}$ is the effective width of the metal tracks.

## D. Study of sensitivity

The distribution of the electric field E at both resonant modes of the reference channel of each device is shown on FIG. 3. It can noted that the E field is more intense in two areas corresponding to the ends of the resonators for the first mode and in an additional third mid-position for the second mode. On the measurement channel of $D_1$, 8 bands of 10

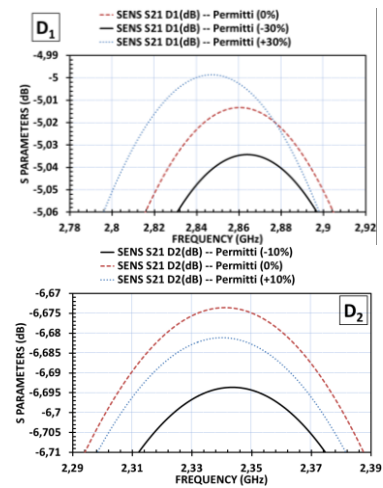

Fig. 4. Simulated influence of the permittivity of the sensitive layer on the transmission parameter ($S_{21}$) of sensitive resonators of $D_1$ and $D_2$.

Full simulations were carried out for the first resonant mode with 10 and 5 sensitive layers respectively for D1 and D2 in order to study the influence of a permittivity variation modeling the absorption of vapor compounds [8]. The results, represented on FIG. 4, show that a permittivity variation of -30% induces for D1 a frequency shift of + 3.75 MHz and -1.25 MHz for a variation of + 30%, according to the direction of the variation in the relation defined in equation (1). Similarly, on D2, a permittivity variation of -10% and + 10% induces a frequency shift of +2.5 MHz and -1 MHz, respectively.

### III. EXPERIMENTAL STUDY

#### A. Manufacture of the sensor

The devices are manufactured using the Dimatix inkjet printer (2800 series), the JSB 25 HV metal ink marketed by the Novacentrix company and a sensitive ink, as the composite polymer solution doped with 1 to 1.2 % of multi-wall carbon nanotubes (PEDOT: PSS-MWCNTs) of reference Poly-Ink HC. The photo paper used as a substrate is Epson brand with a thickness equal to 260 µm. FIG. 5 shows the manufactured devices $D_1$ and $D_2$, with respectively 10 and 5 layers of sensitive material as indicated in part II.

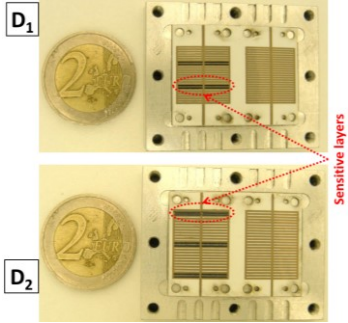

Fig. 5. Devices $D_1$ and $D_2$ manufactured on paper, on the base of the test cell.

#### B. Electrical characterizations

The electrical characterizations are carried out over the frequency range from 1 to 6 GHz with the vector network analyzer (VNA) MS2026B, calibrated with 4001 points. On FIG. 6 are reported the experimental S parameters of each device.

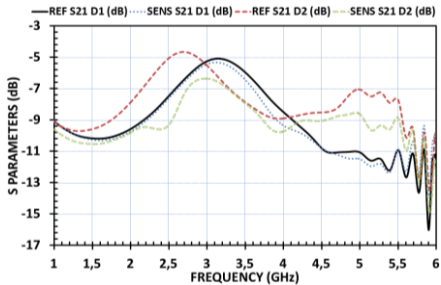

Fig. 6. Transmission parameters ($S_{21}$) of D1 and D2.

We observe that only the first modes can be used over the frequency range considered. The corresponding frequencies are close to those obtained by simulation. Nevertheless, differences related to the deposition of the sensitive material are noted. These differences may be related to the influence of the test and characterization cell of the devices but also to the properties and characteristics of the real materials. Thus, further retro-simulation integrating all parameters would be necessary in order to have a full convergence between simulation and experimental results.

#### C. Characterization with ethanol vapors and sensitivity

The experimental detection configuration is mainly based on a vapor generator (CALIBRAGE PUL 110), which is used to generate and control target vapor concentrations from a liquid contained in a heated bottle. These vapors are transported by nitrogen as a carrier gas at a constant flow equal to 0.112 L/min [9]. A conventional sequence of target vapor concentrations (C) was delivered on the two resonators of a device placed in the low cost dedicated test cell. Ethanol was used as the target vapor, using the sequence: 0, 500, 0, 500, 0, 1000, 0, 1000, 0 and 2000 ppm for 10 minutes each concentration step, carried out after an initial rinsing step under nitrogen for 450 min to ensure stability at room temperature. The relative permittivity of the ethanol is about 24.5 at 25°C, which will cause a change in the environment of the sensor with respect to the permittivity of nitrogen close to 1 and will modify the electrical properties of the sensitive layer [8]. All detections are carried out with the same conditions, at a fixed temperature equal to 26°C and a humidity of 32% RH.

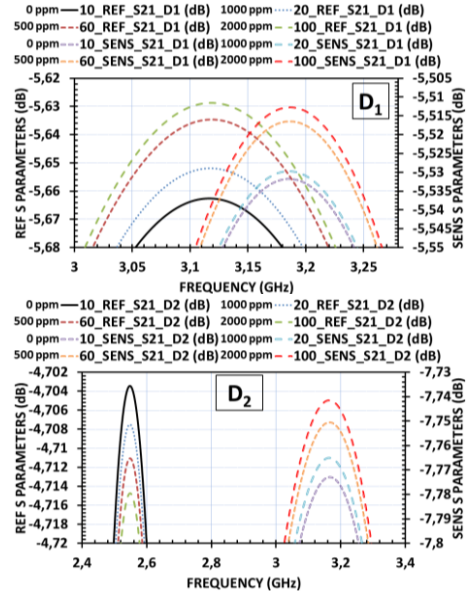

Fig. 7. $S_{21}$ parameters of $D_1$ and $D_2$ for different concentrations of ethanol vapors.

In this section we focus on the behavior of the first mode of the two resonators of $D_1$ and $D_2$ resulting from the interaction of the ethanol vapors with each of them. The so induced modifications in the propagation of the electromagnetic fields are observed in the reduced frequency range between 2 and 4 GHz with the same VNA and 4001 points. FIG. 7 illustrates the effect of each concentration on the $S_{21}$ transmission parameters of the resonators. These measurements correspond to detection in a "static" mode at 10 minutes after the start of the step of exposure, for concentrations 0, 500, 1000 and 2000 ppm of ethanol, respectively at times 10, 20, 60 and 100 minutes of the overall sequence, just before concentration change. The resonance frequencies related to the minimum insertion loss of $S_{21}$ of the reference and sensitive resonators of $D_1$ are noted $FrS_{21r}\_D_1$ and $FrS_{21s}\_D_1$, respectively. They increase slightly as a function of the concentration of ethanol, as shown in Table I. The frequencies of the reference and sensitive resonators of $D_2$ are noted $FrS_{21r}\_D_2$ and $FrS_{21s}\_D_2$ and are reported in an similar manner. In this case, the frequency decreases at the same time as the concentration of ethanol increases, which is in accordance with the expected result, together with an increase of the permittivity of the sensitive material with ethanol vapors.

TABLE I. RESONANT FREQUENCY OF $D_1$ AND $D_2$ AS A FUNCTION OF THE CONCENTRATION OF ETHANOL VAPORS.

| C (ppm) | $FrS_{21r}\_D_1$ (GHz) | $FrS_{21s}\_D_1$ (GHz) | $FrS_{21r}\_D_2$ (GHz) | $FrS_{21s}\_D_2$ (GHz) |
|---|---|---|---|---|
| 0 | 3,1165 | 3,1843 | 2,5490 | 3,1690 |
| 500 | 3,11675 | 3,1845 | 2,5485 | 3,1670 |
| 1000 | 3,1185 | 3,1861 | 2,5480 | 3,1655 |
| 2000 | 3,1190 | 3,1864 | 2,5475 | 3,1630 |

For a better understanding of the response related to the sensitive material under gas, the results are now considered in differential mode, by subtracting the response of the reference line from that of the sensitive one: $\Delta F\_S_{21} = FrS_{21s} - FrS_{21r}$.

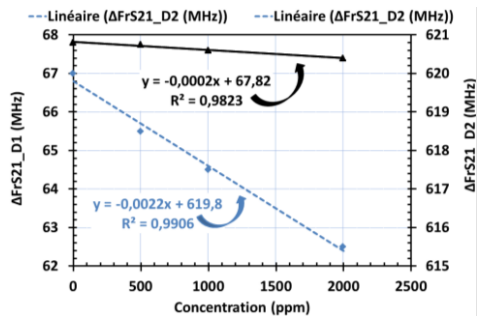

Fig. 8. Estimation of the sensitivity of $D_1$ and $D_2$ for the detection of ethanol vapors.

We thus obtain the responses reported in FIG. 8, which shows the evolution of $\Delta F\_S21$ for the 2 devices after 10 minutes of each concentration step. It is decreasing for both devices, making it possible to verify a coherent behavior, associated with an expected increase of the permittivity of the sensitive material. From this curve, it can be estimated a sensitivity equal to -0.2 kHz/ppm and -2.2 kHz/ppm for $D_1$ and $D_2$, respectively.

## IV. CONCLUSION

The feasibility of the capacitive gas sensor based on electromagnetic transducer has been demonstrated according to the finite element electromagnetic simulations and the experimental study. We have shown that the location of the sensitive layer on the intense E-field zones increases the sensitivity. This can be further improved by increasing the sensitive surface or the thickness of the sensitive material, or by further functionalizing this layer. A more thorough analytical study will improve the understanding of the physical phenomena which are involved. The high active surface area of the proposed structure, which is compatible with a variety of sensitive materials in a large range of electrical conductivity, is of particular interest for the development of various chemical gas sensors based on an integrated microwave device.

ACKNOWLEDGMENT

We would like to thank the National Research Agency for its support (CAMUS project, ANR-13-BS03-0010) and the French Embassy of Singapore (Merlion project), as well as the RENATECH national network for its technological support.